\documentclass[journal]{IEEEtran}

\ifCLASSINFOpdf
\else
   \usepackage[dvips]{graphicx}
\fi
\usepackage{url}

\usepackage{siunitx}
\DeclareSIUnit \belm {Bm} 

\usepackage{graphicx}

\usepackage{soul}

\usepackage[final,nomargin,inline,author=]{fixme}
\fxsetup{theme=color}
\definecolor{fxwarning}{rgb}{0.8,0.0000,0.0000} 
\definecolor{fxnote}{rgb}{0,0.26,0.686} 

\newcommand*\around{{\raise.17ex\hbox{$\scriptstyle\mathtt{\sim}$}}}

\usepackage{textgreek}

\usepackage{changepage}
\usepackage{cite}

\usepackage{lscape}
\usepackage{booktabs}
\usepackage{multirow}
\usepackage[normalem]{ulem}
\useunder{\uline}{\ul}{}
\usepackage{longtable}
\usepackage{tabu}
\usepackage{siunitx}
\DeclareGraphicsExtensions{.png,.pdf,.jpg}

\begin{document}

\title{Programmable Silicon Photonic Optical Thresholder}

\author{Chaoran~Huang,
        Thomas~Ferreira~de~Lima,
        Aashu~Jha,
        Siamak~Abbaslou,
        Alexander~N.~Tait,
        Bhavin~J.~Shastri,
        and Paul~R.~Prucnal
\thanks{This work was supported in part by the Office of Naval Research (ONR) (Award N00014-18-1-2297).}
\thanks{\textsuperscript{*}Corresponding author: chaoranh@princeton.edu}
\thanks{C. R. Huang, T. F. de Lima, A. Jha, S. Abbaslou, A. N. Tait, B. J. Shastri, and P. R. Prucnal are with the Department of Electrical Engineering, Princeton University, Princeton, NJ 08544, USA.}%
\thanks{B. J. Shastri is with the Department of Physics, Engineering Physics \& Astronomy, Queen's University, Kingston, ON KL7 3N6, Canada.}%
\thanks{A. N. Tait is currently with the Physical Measurement Laboratory, National Institute of Standards and Technology, Boulder, CO 80305, USA.}}

\maketitle

\begin{abstract}
We experimentally demonstrate an all-optical programmable thresholder on a silicon photonic circuit. By exploiting the nonlinearities in a resonator-enhanced Mach-Zehnder interferometer (MZI), the proposed optical thresholder can discriminate two optical signals with very similar amplitudes. We experimentally achieve a signal contrast enhancement of 40, which leads to a bit error rate (BER) improvement by 5 orders of magnitude and a receiver sensitivity improvement of 11 dB. We present the thresholding function of our device and validate the function with experimental data. Furthermore, we investigate potential device speed improvement by reducing the carrier lifetime.
\end{abstract}

\begin{IEEEkeywords}
Optical thresholder, nonlinear silicon photonics.
\end{IEEEkeywords}

\IEEEpeerreviewmaketitle

\section{Introduction}

\IEEEPARstart{T}{hresholders} are at the heart of analog-to-digital converters, comparators and operational amplifiers. Thresholders that are based on simple, effective and integrable all-optical components can have operating speeds well beyond the limit of their electronic counterparts. Therefore, all-optical thresholders have found their unique and indispensable role in a variety of applications which require fast signal processing. Examples include neuromorphic photonics, optical code division multiple access (OCDMA), optical logic gate, optical signal regeneration, and physical layer security, etc~\cite{Prucnal:17,kravtsov_simple_2007,bekele_signal_2018,huang2019robustness}. In these applications, all-optical thresholders play a crucial role in effectively enhancing the signal contrast. A poor signal contrast will lead to degradation of the system quality and result in a large bit error rate (BER). An all-optical thresholder can be used to improve the system performance.

Substantial efforts have been made to develop high-performance optical thresholders by exploring different nonlinear effects and materials~\cite{kravtsov_simple_2007,chen_novel_2016,hamanaka_compound_2007,zheng_few-layer_2017}. However, most of these systems are constructed with bulky and discrete photonic devices. Recent work~\cite{bekele_signal_2018} demonstrated an integrated, low-power optical thresholder by exploiting Fano resonances on an InP photonic crystal (PhC) membrane structure. This device improved receiver sensitivity by \SI{2}{\dB}.

The development of the silicon-on-insulator (SOI) platforms offers the possibility to integrate optical thresholders on a silicon chip.
The high refractive index of silicon enables efficient nonlinear interaction of lights within a short waveguide. Moreover, the nonlinearity of silicon can be further enhanced by cavity structures such as microring resonators (MRRs).
Optical thresholders based on silicon MRRs in use with different structures have previously been proposed. However, only numerical simulations have hitherto been shown\cite{tait_dream:_2013,kishikawa_optical_2017}.

In this work, we propose and demonstrate an all-optical thresholder based on resonator-enhanced Mach-Zehnder interferometer (MZI) on a SOI platform, as an extension of our recent conference paper~\cite{huang2019giant}. In addition to signal contrast enhancement demonstration~\cite{huang2019giant}, we present the operation principle of our device using a experimentally-validated theoretical model. Using this model, we also characterize the thresholding function and the processing speed of our device. 

In the ring resonator, near the resonance, the signal experiences a power-dependent nonlinear phase shift that varies rapidly with its optical power. In addition, the ring resonator can significantly enhance the circulating power of a signal. Our thresholder makes use of the power-dependent nonlinear phase, and the MZI can convert such phase shift to amplitude change with a large extinction ratio~\cite{heebner1999enhanced}. The combined action of these effects leads to a highly sensitive thresholder with a sharp power transfer slope of 44. In addition, we have implemented a Mach-Zehnder coupler (MZC) preceding the MZI together with thermally tuned biases on MZC, MZI and ring resonator. Such a design makes this device fully programmable, therefore ensuring that perfect interference and hence largest sensitivity can be achieved. With the proposed thresholder, we experimentally demonstrate that two optical signals with very close power levels can be well distinguished, leading to ~40x signal contrast improvement. This consequently leads to a BER improvement by 5 orders of magnitude and a receiver sensitivity improvement of \SI{11}{\dB}.

In a silicon waveguide, both the Kerr effect and the free carrier dispersion (FCD) can induce a nonlinear phase shift on traveling lights with fast dynamics. It is found that FCD practically dominates over Kerr in the silicon MRRs~\cite{de2019enhancing}. Therefore, in our current device, FCD is the dominating mechanism that contributes to the nonlinear phase and results in the thresholding function. FCD, stemming from two-photon absorption (TPA) induced carriers, has a long response time in the order of hundreds of picoseconds, and thus slow down the device speed. Therefore, we propose to improve the device speed through active carrier removal, i.e., reverse-biasing a p-i-n junction transversal to the silicon waveguide to sweep away the carriers~\cite{rong_all-silicon_2005,liang2004nonlinear}. We show that the proposed device can operate at 10 GHz when the carrier lifetime is reduced to $\sim$ 18 ps. The carriers imposed speed limitation can also be overcome by other techniques, including the use of a silicon-organic hybrid approach and other TPA-free nonlinear materials in which the Kerr effect becomes the dominated mechanism contributing to the phase shift~\cite{leuthold_silicon_2009,moss_new_2013}. All these techniques are compatible with our proposed thresholder.

\section{Device Principle, Design and Characterization}

\begin{figure}
\centering\includegraphics[width= 85mm]{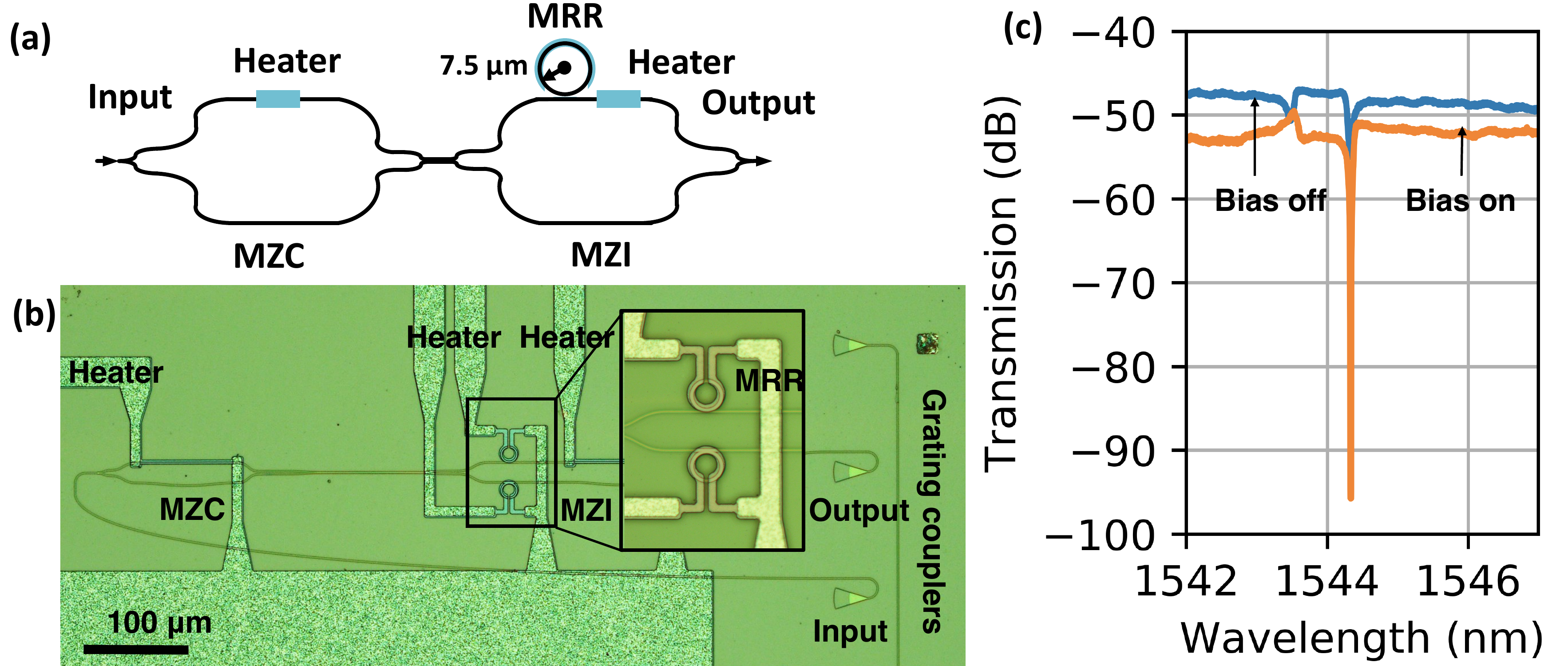}
\caption{(top-left) Schematic illustration of the proposed all-optical thresholder; (bottom-left) Microscope image of the fabricated device. The resonance of the MRR on the lower arm of MZI is tuned away from the operating wavelength. (right) Optical spectrum of the MZI before and after tuning the heaters to the desired regime of operation.}
\label{new_schematic diagram}
\end{figure}

The idea of the proposed optical thresholder is to exploit the optical-power-dependent phase shift induced by the nonlinearity in silicon waveguide. An MZI is used to convert the phase change into an intensity change with a large extinction ratio. With a sufficiently large phase difference due to the power-dependent nonlinearity, the interference between the signals from the two arms of the MZI can switch from constructive to destructive, leading to self-switching. To achieve a significant self-switching under low optical powers, we load an MRR in one arm of the MZI, resulting in an all-optical thresholder based on an MRR-enhanced MZI (Fig.~\ref{new_schematic diagram}(a)). This enhancement considerably reduces the required optical power supply by increasing the effective interaction length and instantaneous optical power through coherent power buildup.

To maximize the thresholding effect, it is critical to switch off the low power signal through destructive interference. Perfect destructive interference requires the signals traveling in the two MZI arms to have equal amplitudes and an exact \textpi~phase difference. Therefore, we designed a MZC preceding the MRR-loaded MZI through a wideband \SI{3}{\dB} coupler. The bias of the MZC (through the heater on one of its arms) can be adjusted to balance the amplitudes at the two arms of the MZI, while the MZI bias can be independently tuned to introduce a \textpi~phase difference. 
The bias on the MRR also needs to be carefully adjusted to ensure that the thresholder is working around the resonance wavelength to achieve the highest sensitivity.

Our all-optical thresholder consists of fully-etched, \SI{500}{\nm}-wide waveguides (Fig.~\ref{new_schematic diagram}(b)) on a passive-SOI platform with silicon thickness of 220 nm, a \SI{3}{\um} oxide passivation layer, a Ti/W heating filament layer, and an Al routing layer. The MRR on the MZI's arm has a radius of \SI{5}{\um} and high coupling coefficient (gap = 100 nm), yielding a Q-factor \around 25000. The resonance of the MRR on the lower arm of MZI is tuned away from the operating wavelength and is not used in our experiment. A microheater on the MRR provides flexible resonance control over a full free spectral range (FSR). Thus, input signals of different wavelengths can be easily accommodated. Two microheaters are deposited on the arms of MZC and MZI. These tunable elements can control the interference condition of the device and enable us to locate the sweet spot of thresholding for the signals.

Two typical transmission spectra under different microheater DC current biases are shown in Fig.~\ref{new_schematic diagram}(c). When the biases are off (blue curve), the resonance features on the transmission spectrum resemble a Lorentzian-like shape with an on-off ratio of \around \SI{7.5}{\dB}. However, there is a slight asymmetry in the shape due to the residual path unbalance between the two MZI arms. When the bias currents are on and adjusted (orange curve), the optical power at resonance is \around -90 dBm denoting an off condition. The on-off ratio in this case is found to be more than \SI{45}{\dB}. This result indicates that, loading an MRR on the MZI can significantly improve the on-off ratio of the device transfer function. This highly sensitive transfer function can be explained by the Fano resonance effect, which results from the interference between a resonance pathway (MRR) and a coherent background pathway (MZI)~\cite{zhou_fano_2007,limonov_fano_2017}.

\section{Signal Contrast Enhancement Experiment and Results}\label{Experiment_Setup_and_Results}

\begin{figure}
\centering\includegraphics[width=0.75\linewidth]{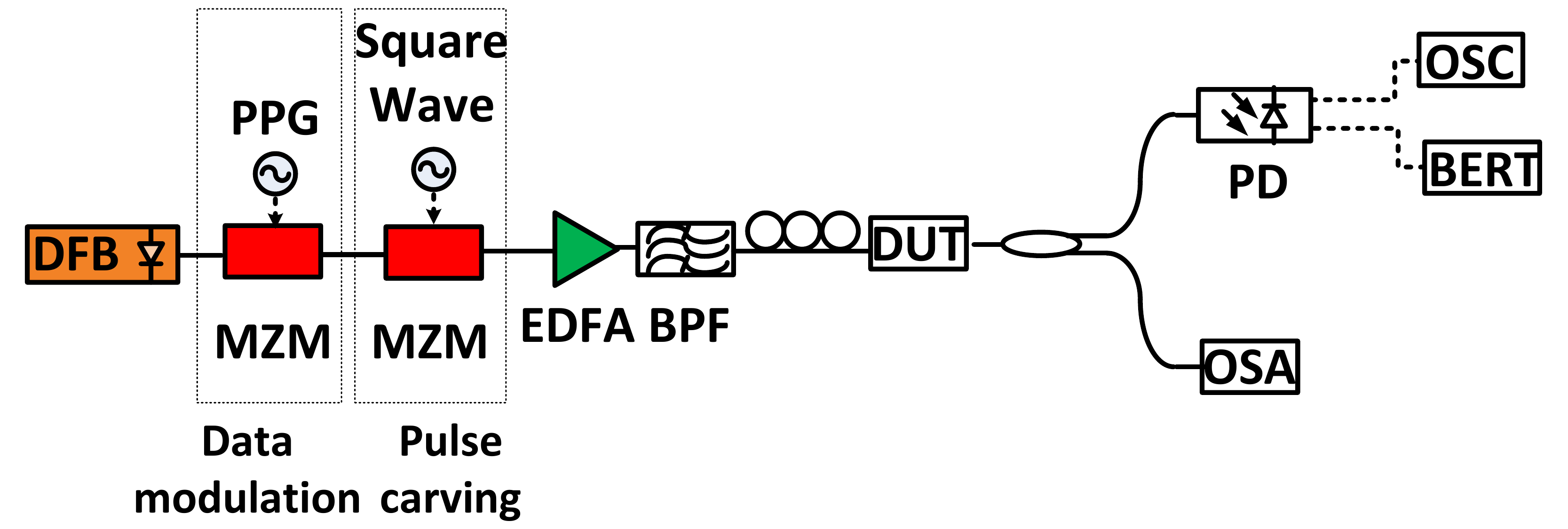}
\caption{Experimental Setup: DFB: distributed feedback laser; MZM: Mach-Zehnder; PPG: pulse pattern generation; EDFA: erbium-doped fiber amplifier; BPF: bandpass filter; DUT: device under test; OSA: optical spectrum analyzer; OSC: oscilloscope; BERT: bit error rate tester.
}\label{experimental_setup}
\end{figure}

The experimental setup is shown in Fig.~\ref{experimental_setup}. The signal is generated by modulating a distributed feedback (DFB) laser output using two cascaded MZMs. The first MZM is driven by electrical pulses from a pulse pattern generator (PPG) (Anritsu MP1763b). A pulsed optical signal with \around \SI{80}{\ps} pulsewidth and equalized peak power is generated. The second MZM is driven by programed patterns at a data rate of \SI{400}{\mega b\per\second}. This yields a \SI{400}{\mega b\per\second} return-to-zero (RZ) signal with two different power levels, and the contrast between two power levels can be dynamically adjusted by tuning the bias of the second MZM. The data speed is limited by the decay time of the TPA-induced carriers. The optical signal is amplified to \SI{20}{\deci\belm} by an erbium doped fiber amplifier (EDFA) to trigger the nonlinearity in the silicon waveguide and compensate for the fiber-to-chip coupling loss. The optical signal is coupled to the device by free-space coupling through a sub-wavelength grating coupler with \around \SI{8}{\dB} coupling loss. The eye diagrams of the input and output signals are obtained by photodetectors and monitored using a sampling oscilloscope (OSC) (Tektronix DSA8300). The signal optical spectrum is monitored using an optical spectral analyzer (OSA) (APEX AP2440A). The microheaters are independently driven by computer-controlled current sources to optimize the parameters necessary to attain a high signal contrast ratio.

\begin{figure}
\centering\includegraphics[width=1\linewidth]{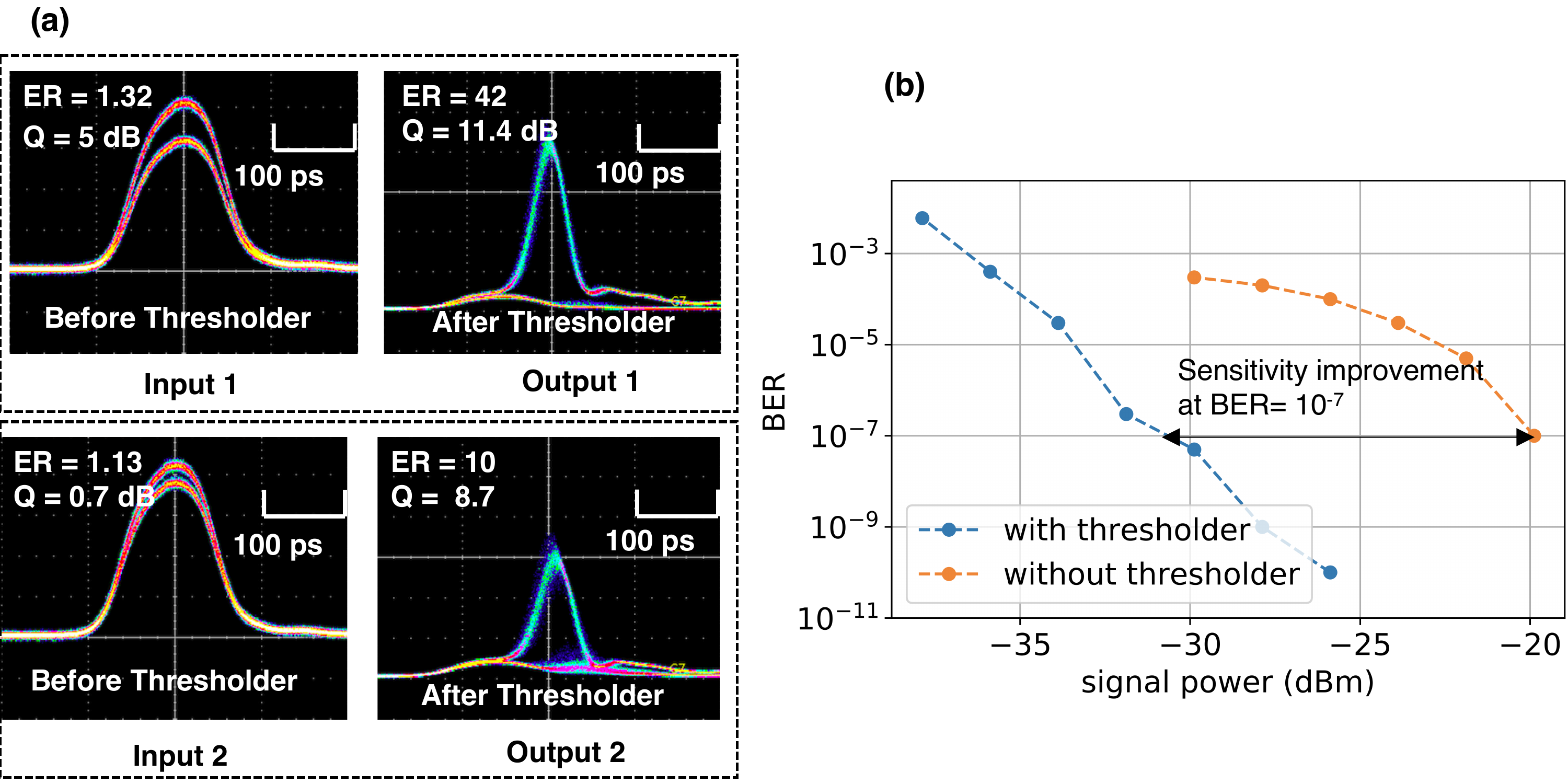}
\caption{Eye diagrams of the signals before (left) and after (right) the thresholder for input signal contrast ratios of 1.32 (top) and 1.12 (bottom).
}\label{eye_diagram_BER}
\end{figure}

Fig.~\ref{eye_diagram_BER}(a) shows the device performance using two sets of signals with different input signal contrast ratios. Both sets of signals have contrast ratios close to 1, resulting in significantly degraded signal quality (Q-factor) even though the received average powers (\SI{0}{\deci\belm}) are much higher than the receiver sensitivity. After being processed by the thresholder, the lower power pulses in both signals are fully suppressed. As a result, the signals after thresholding have a significant signal contrast enhancement (\around 40~times for signal 1, and 7.5~times for signal 2), which leads to a Q-factor improvement of \SI{6.4}{\dB} for signal 1 and \SI{8}{\dB} for signal 2. The result confirms that our thresholder works well under signal contrast close to 1.

Fig.~\ref{eye_diagram_BER}(b) shows the results of BER measurement of signal 2 using a BER tester (BERT). Assisted with the all-optical thresholder, the communication link can achieve an error-free detection (BER = $10^{-9}$) at a received signal power of \SI{-27.5}{\deci\belm} due to the contrast enhancement leading to an opened eye. \fxwarning{need to revise the experiment setup figure to explain the received power} Without the thresholder, at the same received power (\SI{-27.5}{\deci\belm}), the link has a BER higher than $10^{-4}$. The presence of this thresholder can also effectively improve the receiver sensitivity by 11 dB at a BER of $10^{-
7}$(see Fig.~\ref{eye_diagram_BER}(b).


\section{Thresholding tranfer function}\label{Thresholding_tranfer_function}

To correctly model the thresholding behavior of our device, nonlinearities in the silicon waveguide including the Kerr effect, TPA, TPA induced free-carrier absorption (FCA) and FCD are taken into consideration. Thermal-optic effect is excluded due to its long response time compared to the signal speed. In our simulation model, the MZC and MZI are treated as linear waveguides due to their short lengths. Nonlinear coupled-mode theory is used to study the change in the signal complex amplitude and carrier density in the MRR~\cite{chen_bistability_2012}. The evolution of the normalized complex amplitude $a$, and the normalized carrier density $n$ is governed by 
\begin{IEEEeqnarray}{rCl}
\IEEEyesnumber\label{eq:ode}
\partial a/\partial t &=& i(\delta\omega - n_\mathrm{Kerr}|a|^2 + \sigma_\mathrm{fcd}\alpha_\mathrm{tpa}n)a \,- \nonumber\\
&&-(1 + \alpha_\mathrm{tpa}|a|^2+\gamma_\mathrm{fca}\alpha_\mathrm{tpa}n)a + \sqrt{\gamma_p P_\mathrm{in}(t)}\IEEEyessubnumber\label{eq:cavity}\\
\partial n/\partial t &=& |a|^4 - n/\tau,\IEEEyessubnumber\label{eq:carriers}
\end{IEEEeqnarray}
where $\delta\omega$ is the frequency detuning between the light source and the MRR resonance; $t$ is the time variable normalized with $\Gamma_0^{-1} = 2Q_L/\omega_0$, $Q_L$ is the total quality factor; $P_\mathrm{in}$ is the power input, and $(n_\mathrm{Kerr}, \alpha_\mathrm{tpa}, \sigma_\mathrm{fcd}, \gamma_\mathrm{fca}, \gamma_p) \propto (n_2\omega_0, \beta_2, \sigma_{e,h}\omega_0, \sigma_\mathrm{fca}, \Gamma_c/\Gamma_0^3)$, are the Kerr, TPA, FCD, FCA, and quality factor coefficients, respectively. These equations were simplified from~Ref.~\cite{chen_bistability_2012}, and renormalized so that the two-photon absorption term only appears in~Eq.~(\ref{eq:cavity})~\cite{de2019enhancing}. 

\begin{figure}
\centering\includegraphics[width= 65mm]{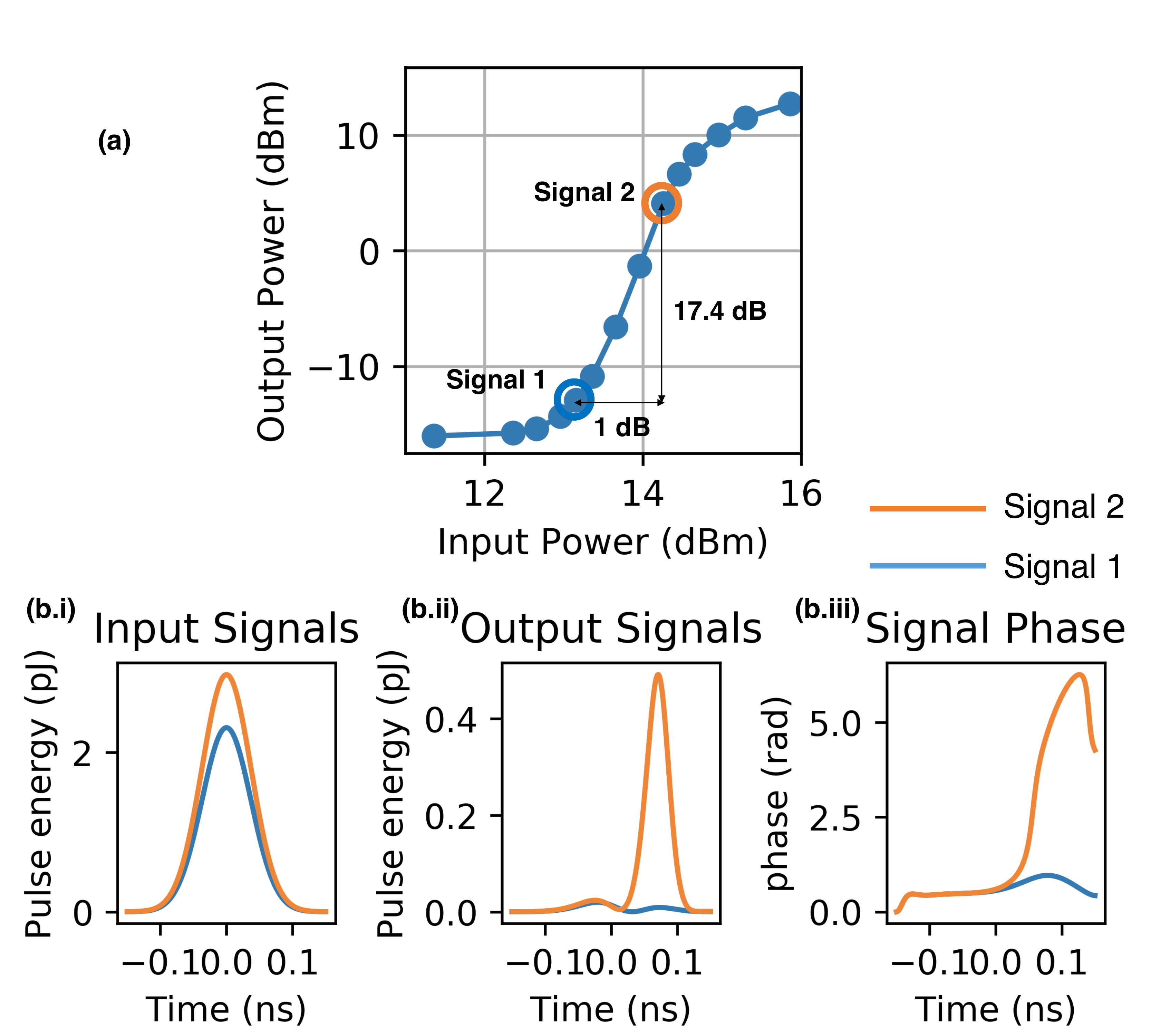}
\caption{(a) Power transfer function of the proposed device; (b) Input (i) and output (ii) signal waveforms with two different peak powers; (iii) signal phase evolution.}\label{thresholding_function}
\end{figure}

The input signals are Gaussian pulses with widths of 100 ps. Their wavelength is located at 150 GHz away from the MRR resonance, and the MRR Q factor is 25000. These conditions are consistent with those in the experimental measurement. 
The power splitting ratio on MZC and the phase bias on MZI are optimized such that the slope of the transfer function is maximized. The transfer function in Fig.~\ref{thresholding_function}(a) shows that, through our thresholder, a signal contrast of 1 dB (the signal contrast is numerically equal to the extinction ratio (ER) of 1.25) between the two input signals is enhanced to 17.4 dB (ER = 54.9) in the output signals, resulting a 44$\times$ signal contrast enhancement. Fig.~\ref{thresholding_function}(b.iii) shows the nonlinearity-induced intensity
dependent phase change in the MRR, which renders an amplitude shift in the MZI output. Along with this phase change, we
can optimize the biases applied to the MZC and MZI to maximize the ratio of the peak powers between two output signals. This
can be accomplished when the phase difference of the “1”-level signal and “0”-level signal is approximately \textpi, and a destructive interference occurs on the “0”-level signal while a constructive interference occurs on the “1”-level signal. In the simulation, we also verify that FCD significantly dominates over the Kerr effect and governs the nonlinear phase shift. Such a verification is conducted by observing the difference in signal phase when the Kerr effect is included and excluded, and we find that the phase difference is insignificant (the results are not shown in this paper) in these two cases. The input and output pulses of the two signals are plotted in Fig.~\ref{thresholding_function}(b.i) and (b.ii). The simulated contrast enhancement and the pulse waveform match well with the experimental data.

\section{Device speed discussion}\label{Processing_bandwidth_discussion}

Although FCD plays a dominant role in discriminating the signals, its long lifetime hinders fast nonlinear signal processing ($>$10~GHz) in silicon. Therefore, the processing speed of the current device is limited to 400 Mbit/s. A widely applied technique to overcome the speed limitation is by active carrier removal, i.e., reverse-biasing a p-i-n junction transversal to the silicon waveguide to reduce the lifetime of free carriers. The carrier lifetime can be effectively reduced by increasing the reverse-biasing voltage~\cite{turner2010ultrashort}. 

\begin{figure}
\centering\includegraphics[width= 50mm]{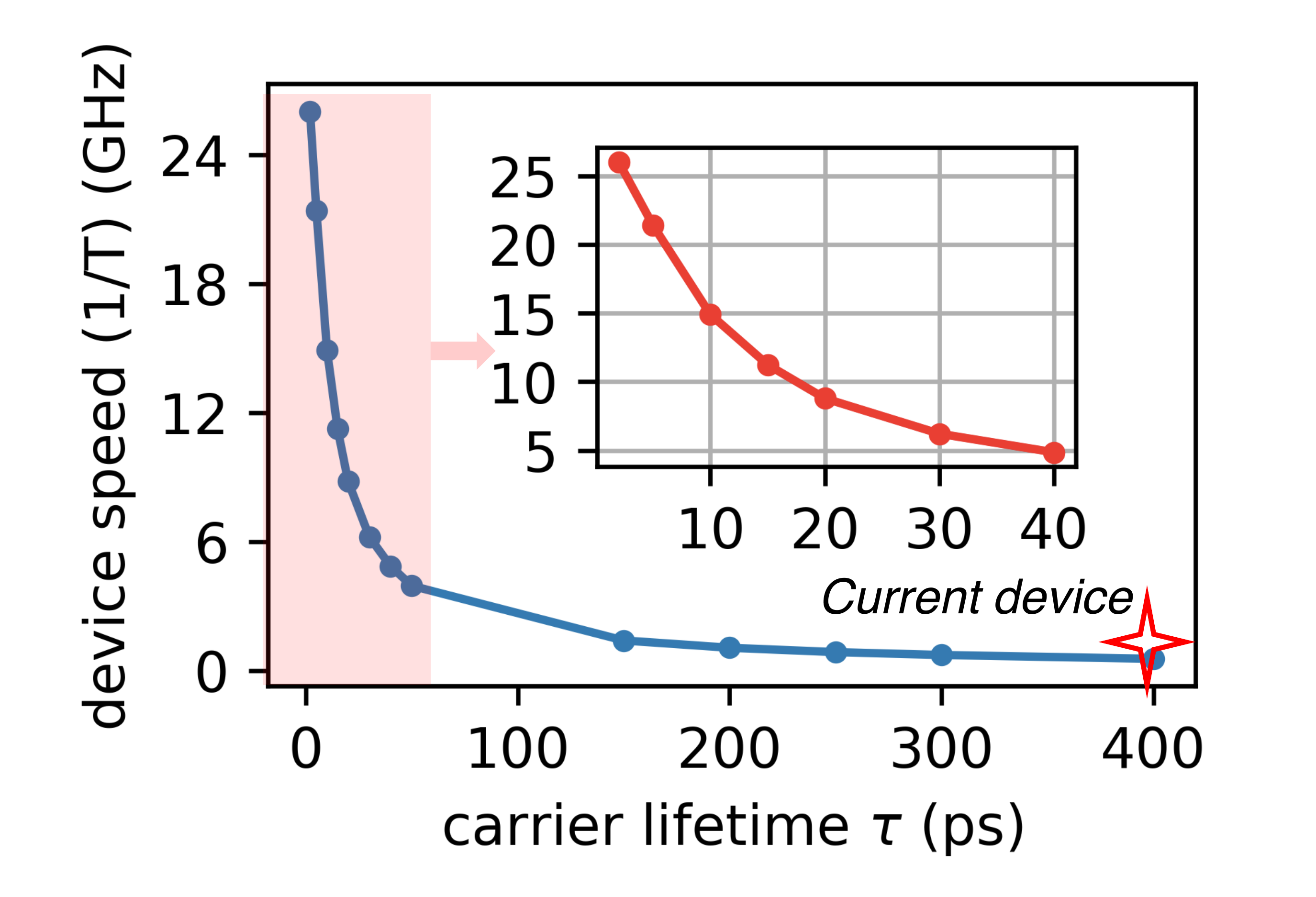}
\caption{Device speed as a function of the carrier lifetime; inset: zoom-in view of red region.}\label{carrier_lifetime}
\end{figure}

Here, we study the device speed with active carrier removal and characterize the device speed under different carrier lifetime $\tau$ using our simulation model described in Eq.~\ref{eq:ode}. In device speed characterization, the input signal is an impulse with a pulsewidth $<$1~ps. The device speed is defined as $1/T$, where $T$ is the time that takes to reduce the free carrier number by 99\% compared to the peak carrier number. It is worth noting that the definition of $T$ here takes the cavity effect of MRR into consideration, and thus is not equivalent to the carrier lifetime. Fig.~\ref{carrier_lifetime} shows the device speed as a function of the carrier lifetime. As expected, reducing the carrier lifetime can increase the device speed. With reduced lifetime, the similar thresholding function can still be achieved at the cost of requiring a higher signal power. Our current device operates at a speed of 400 Mbit/s and is marked in Fig.~\ref{carrier_lifetime}. The inset of Fig.~\ref{carrier_lifetime} is a zoom-in view when the carrier lifetime is smaller than 40~ps. As shown in the inset, our thresholder has the potential of working beyond 10 GHz when the carrier lifetime is reduced to $\sim$~18~ps~\cite{turner2010ultrashort}. 


The processing speed limitation imposed by carrier effects can be further relaxed by designing MRR with a lower Q factor. Other alternative approaches include the use of a silicon-organic hybrid waveguide and other TPA-free nonlinear materials~\cite{leuthold_silicon_2009, moss_new_2013}. All these methods are compatible with the design of our proposed thresholder.

\section{Conclusion}\label{Conclusion}

We have proposed and experimentally demonstrated an all-optical programmable nonlinear thresholder based on resonator-enhanced nonlinearity in a Mach-Zehnder interferometer. This device can discriminate signals
with extremely close power levels due to its sharp thresholding transfer function, as predicted by a theoretical model based on nonlinear coupled theory. We experimentally demonstrated that this thresholder enables an enhancement of 40 times in signal amplitude contrast, and consequently, an improvement of 11 dB in the receiver sensitivity. The proposed thresholder, developed on a CMOS-compatible silicon-on-insulator (SOI) platform, can find uses in a number of high-performance optical signal processing applications and can be monolithically integrated with other on-chip functionalities.

\bibliographystyle{IEEEtran}

\bibliography{2019_PTL}

\end{document}